\documentclass[reprint,aps,prab,amsmath,amssymb,nofootinbib,floatfix,superscriptaddress,longbibliography]{revtex4-1}
\usepackage[utf8]{inputenc}
\usepackage{graphicx}
\usepackage{hyperref}

\begin{document}
\title{Demonstration of Two-Color X-ray Free-Electron Laser Pulses with a Sextupole Magnet}

\author{P.~Dijkstal}
\email[]{philipp.dijkstal@psi.ch}
\affiliation{Paul Scherrer Institut, 5232 Villigen PSI, Switzerland}
\affiliation{Department of Physics, ETH Zürich, 8092 Zürich, Switzerland}

\author{A.~Malyzhenkov}
\affiliation{Paul Scherrer Institut, 5232 Villigen PSI, Switzerland}

\author{S.~Reiche}
\affiliation{Paul Scherrer Institut, 5232 Villigen PSI, Switzerland}

\author{E.~Prat}
\email[]{eduard.prat@psi.ch}
\affiliation{Paul Scherrer Institut, 5232 Villigen PSI, Switzerland}

\date{\today}

\begin{abstract}
We present measurements of two-color X-ray free electron laser (FEL) pulses generated with a novel scheme utilizing a sextupole magnet.
The sextupole, in combination with a standard orbit control tool, is used to suppress the radiation from the bunch core, while keeping the head and the tail of the beam lasing, each at a different photon energy.
The method is simple, cost-effective and applicable at any repetition rate.
We demonstrate the tunability of the scheme and discuss its advantages and practical limitations.
\end{abstract}

\maketitle

X-ray free-electron lasers (FEL) emit radiation at very short wavelengths down to the angstrom level.
The pulses can reach peak powers of several tens of gigawatts and typically have durations between a few and 100~fs~\cite{Pellegrini2016,Seddon2017}.
The light generated at X-ray facilities is used to advance research areas such as physics, chemistry and biology (see for example~\cite{Ullrich2012,Spence2012,Kern2015,Bostedt2016}).
There is special scientific interest in two-color FEL pulses with time delay and central photon energy separation both adjustable.

Among the use cases are pump-probe~\cite{Inoue2016,Ferrari2017a,Pontius2018,Lu2018,Bencivenga2019} experiments, coherent stimulated X-ray Raman spectroscopy~\cite{Schweigert2007,Harbola2009,Weninger2013,Picon2015}, multi-wavelength anomalous diffraction~\cite{Gorel2017a}, time-resolved coherent diffraction imaging~\cite{Zangrando2013}, and static two-color imaging~\cite{Weder2017}.

The FEL wavelength resonance condition for planar undulators is a function of the electron beam Lorentz factor $\gamma$, the undulator period length $\lambda_u$, and the undulator field strength parameter $K$~\cite{Bonifacio1984}:
\begin{align}
	\lambda = \frac{\lambda_u}{2\gamma^2} \left(1+\frac{K^2}{2}\right)
	\label{eq:resonance}
\end{align}
The photon energy is given by $E={hc}/{\lambda}$, with $h$ the Planck constant and $c$ the speed of light.
Since the undulator period is fixed for a given beamline, there are generally two possibilities for two-color pulse generation: a change of $K$ (provided the undulator gap can be varied) or of $\gamma$.

In the first case, a single electron bunch is lasing in two subsequent undulator sections with different undulator strengths~\cite{Jaroszynski1994,Lutman2013,Hara2013}.
In this scheme the power of the two FEL pulses is limited, since the same bunch is used to generate FEL radiation in the two stages.
This disadvantage can be overcome by allowing only a part of the beam to lase at each section.
For instance a beam tilt~\cite{Reiche2016,Lutman2016,Guetg2018} (i.e., a correlation between the transverse and longitudinal positions of the electrons) or a longitudinal slice-dependent optics mismatch~\cite{Qin2017,Chao2018} can be used to this end.
The first color is produced with only a part of the bunch, whereas the second color is emitted by another, still unspoiled part.
A drawback of these methods is the need for a long undulator beamline with two sections.
An advantage is the large and independent tunability of two important properties: the temporal separation of the two pulses can be varied widely with a delaying chicane between the two undulator sections, and the energy separation can be tuned by adjusting the undulator strength of each section.

A second option to obtain two-color FEL pulses employs two separated ensembles of electrons at different energies with all undulators tuned at the same strength.
Compared to the first type of schemes, this approach has the advantage of being compact, since only one undulator section is required.
It was first realized with two bunches at different energies
~\cite{Marinelli2015,Marinelli2016a,Decker2017}.
Several methods have been proposed to achieve the same with a single energy-chirped bunch.
They all share the necessity to suppress lasing from the central parts of the bunch, using two current spikes and an otherwise low current~\cite{Bettoni2016},
a longitudinally shaped laser heater pulse~\cite{Marinelli2016}
, or a slotted foil as an emittance spoiler~\cite{Emma2004,Ding2015,Saa2019}.
A downside of the spoiler technique is that it generates significant radiation losses, which may limit the repetition rate of the accelerator and hence the produced X-ray pulses.

In this paper we present experimental results of a novel, simple and cost-effective method using a single bunch~\cite{Dijkstal2019}.
The demonstration has been carried out at SwissFEL~\cite{Milne2017a} at the Paul Scherrer Institute in Switzerland.
We first explain our method in detail before briefly describing the SwissFEL facility.
We then show measurements of electron beam and photon pulse properties.

Our method of two-color generation makes use of the fact that the FEL process requires sufficient transverse overlap between the electron distribution and the radiation field.
Off-axis bunch slices lose transverse overlap due to kicks from the quadrupoles that are periodically placed between the undulator modules.
The threshold of orbit misplacement, above which the FEL process is suppressed, depends on the photon energy and on electron beam parameters such as the transverse size~\cite{Emma1999,Tanaka2004,MacArthur2018}.
Our simulations show that for SwissFEL parameters and a photon energy of 12~keV, a root-mean-square trajectory misplacement of about 20~$\mu$m is required for a suppression of the FEL power by a factor 5.
To generate two pulses separated in time, we employ a quadratically tilted beam to obtain a suitable correlation of the trajectory with the slice position, for which the bunch tails propagate on axis and the central part off axis.

Tuning a sextupole magnet in a dispersive bunch compressor of the FEL facility is an elegant way to impose a quadratic beam tilt.
In the following we assume a bunch compressor acting in the horizontal plane, although the principle works in either transverse plane.
The beam is necessarily energy-chirped while undergoing compression.
When the bunch core is at an energy equal to the chicane reference energy, it travels on axis, whereas the other slices, with different energies, are offset in the horizontal plane.
A sextupole field has a quadratic dependence on the horizontal position, therefore it kicks both bunch tails to one side leaving the bunch core unaffected.
The result is a kick with strength depending quadratically on the longitudinal particle position, generating the desired second-order beam tilt.

\begin{figure}[h]
	\centering
	\includegraphics[width=0.31\textwidth]{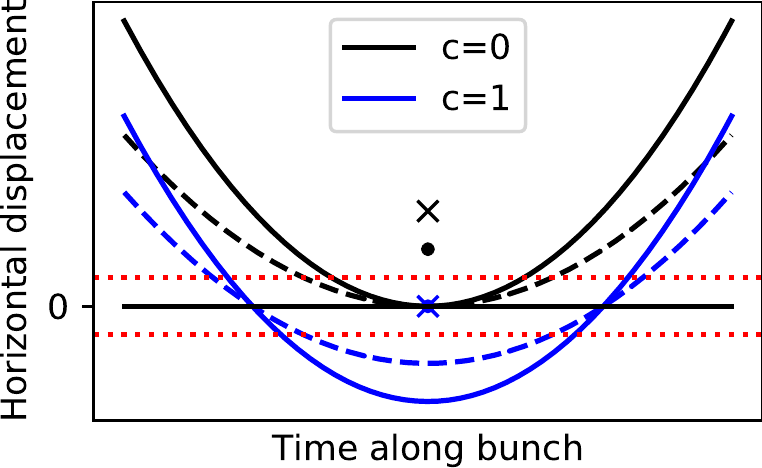}
	\caption{
		Schematic showing the effect of a sextupole magnet on the slice alignment at a location downstream of the sextupole.
		Dashed lines correspond to a weaker sextupole strength.
		The crosses (strong sextupole field) and dots (weak) depict the center of mass.
		For positive values of the orbit adjustment factor $c$, two colors can be generated.
		The threshold of orbit misalignment necessary to suppress FEL radiation is indicated in red.
	}
	\label{fig:scheme_explain}
\end{figure}

If the whole bunch lases when the sextupole is turned off, only the central part keeps lasing after switching it on.
We add a global transverse shift to the electron beam to obtain two separate regions that are well aligned to the nominal trajectory.
For this purpose we employ a standard orbit control tool using beam-position monitors (BPM) and dipole corrector magnets.
The BPMs measure the centroid variations induced by the sextupole magnet along the accelerator.
An adjustment of the amplitude of these variations allows us to align different parts of the bunch.
For a measurement of the initial orbit deviation we record the BPM readings when the sextupole is off ($\mathbf{x}_1$) and on ($\mathbf{x}_2$).
After that we impose a new orbit: $\mathbf{x}_3 = \mathbf{x}_2 + c\cdot (\mathbf{x}_1 - \mathbf{x}_2)$, where $c$ is an adjustment factor.
For $c=0$ the bunch core remains aligned and lases.
For $c>0$ two colors can be generated.
The effect is illustrated schematically in Fig.~\ref{fig:scheme_explain}.

The time separation between the two pulses is adjusted by tweaking $c$.
The bunch duration is an upper limit of the separation, while the requirement that lasing from the central part is suppressed imposes a lower limit.
The individual pulse duration depends on the local slope ($dx/dt$) of the aligned slice (shorter pulse for a larger slope).
The local slope increases linearly with the sextupole strength and with the distance of the aligned slice from the bunch center (see Fig.~\ref{fig:scheme_explain}).
The temporal separation and the individual pulse duration are therefore anticorrelated for constant sextupole strength.
The longitudinal dispersion of a dispersive section could be used to further improve the tunability of the method~\cite{Saa2019}.

\begin{figure*}
	\centering
	\includegraphics[width=1.0\textwidth]{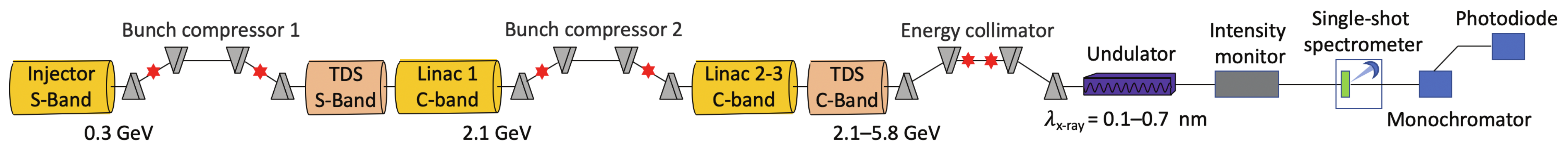}
	\caption{
		SwissFEL schematic (not to scale).
		Six sextupole magnets (marked in red) placed in the dispersive sections are available for our two-color scheme.
		Two transverse-deflecting cavities (TDS) can streak the electron beam in the vertical direction for longitudinal diagnostics.
	}
	\label{fig:swissfel}
\end{figure*}

When the head and the tail of the bunch exhibit different energies while passing the undulator section, they emit radiation at different central wavelengths according to Eq.~\ref{eq:resonance}.
The energy chirp can be obtained with off-crest acceleration or wakefield dechirper modules~\cite{Bane2012}.
For a given time separation, the photon energy separation of the two pulses can be varied independently by adjusting the energy chirp.

The performance of the scheme is limited by three effects of the sextupole magnet that are detrimental to the electron beam quality.
First, the slice horizontal beam emittance increases, scaling with the third power of the beam size and linearly with the sextupole strength~\cite{Dowell2018}.
Second, beam chromaticity is generated, which translates to a slice optics mismatch~\cite{Minty2003}.
Third, the sextupole kick changes the path length of the particles through the bunch compressor chicane depending on their energy, causing a current profile skewness along the bunch.
The first two effects can be mitigated with a small $\beta$ function at the location of the sextupole~\cite{Dijkstal2019}.
Moreover, the chromaticity can be corrected without changing the tilt by using additional quadrupoles and sextupoles~\cite{Guetg2015}.

We report on the demonstration of our method at the SwissFEL facility, schematically shown in Fig.~\ref{fig:swissfel}.
Electrons originate from an rf photocathode with a repetition rate of up to 100~Hz and are accelerated to an energy of up to 5.8~GeV.
Two bunch compressors (BC1 and BC2), acting in the horizontal plane, shorten the bunch length.
Two normal quadrupole, two skew quadrupole and two sextupole corrector magnets are located in each bunch compressor and in the energy collimator (EC).
In standard operation, these magnets correct first- and second-order beam tilts~\cite{Guetg2015}.
Any of the six available sextupoles can impose the beam tilt required for two-pulse generation, as discussed above.
Two transverse-deflecting structures (TDS)~\cite{Craievich2013}, streaking in the vertical plane, can be used in combination with a profile monitor~\cite{Ischebeck2015} to measure longitudinal electron properties such as bunch length and current profile, as well as horizontal slice properties such as slice emittance, beam tilt and optics mismatch~\cite{Prat2019}.
The undulator beamline contains 13 planar variable-gap undulator modules with a period of 15~mm.
Photon pulse energies and spectra are measured by a gas detector and a single-shot photon spectrometer (PSSS), respectively~\cite{Juranic2018}.
Moreover, a monochromator followed by a photodiode is suitable for measuring the average spectral intensity over a large photon energy range.
For the moment, in the absence of direct measurements of the photon pulse in the time domain, we rely on energy-chirped electron bunches to relate spectral and temporal properties of the photon pulse.

To verify our method of beam tilt generation, we can measure the beam after BC1 or BC2 using a TDS.
The betatron phase advance is varied by up to 180$^\circ$ with a quadrupole scan~\cite{Wiedemann2015}.
At every step, an image taken from a beam screen is divided into vertical slices.
For each slice, the centroid positions are determined with respect to a reference slice.
Using the linear beam transport formalism~\cite{Wiedemann2015}, we obtain the slice centroid position and angle at a reconstruction point upstream of all varied quadrupoles.

\begin{figure}
	\centering
	\includegraphics[width=0.49\textwidth]{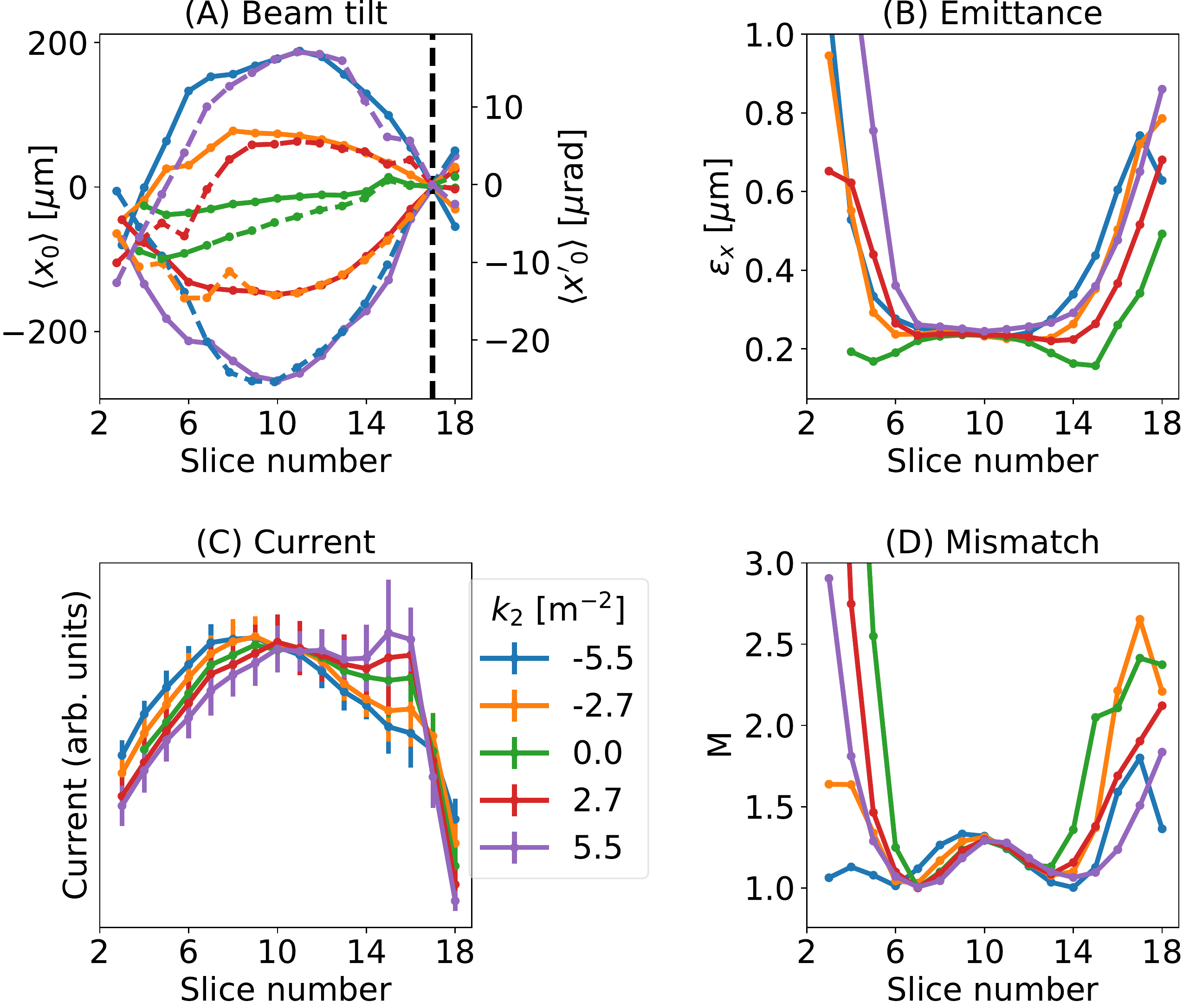}
	\caption{
		Effect of a sextupole magnet on the electron beam properties.
		The strength ($k_2$) of the first sextupole in BC1 was varied in five steps in the available range.
		The beam was streaked with the TDS after BC1.
		A: reconstructed slice position and angle (dashed), with slice 17 as reference.
		B and C: measured slice emittance and slice current.
		D: slice optics mismatch parameter calculated with respect to the design optics.
	}
	\label{fig:m_scan}
\end{figure}

Figure~\ref{fig:m_scan} shows measurements of the beam tilt, slice emittance, current, and optics mismatch for different strengths of the first sextupole in BC1.
As expected, we observe a second-order beam tilt suitable for the generation of two-color FEL pulses.
(There is however a visible asymmetry as a consequence of an initial linear tilt.
Such an asymmetry would prevent a simultaneous exact alignment of two opposing slices in both centroid angle and position, causing also an asymmetry in the observed FEL spectrum.
It could be corrected by imposing a linear tilt of opposite sign in both angle and offset, using for instance two quadrupoles in a dispersive section.)
The expected change in the current profile due to the sextupole field is observed: current is shifted towards one side of the bunch, depending on the sign of the sextupole polarity.
Very little mismatch and emittance increase develop in the central parts of the bunch.
In contrast, emittance and optics mismatch of the outer slices are strongly affected by large sextupole strengths.

\begin{figure}
	\centering
	\includegraphics[width=0.49\textwidth]{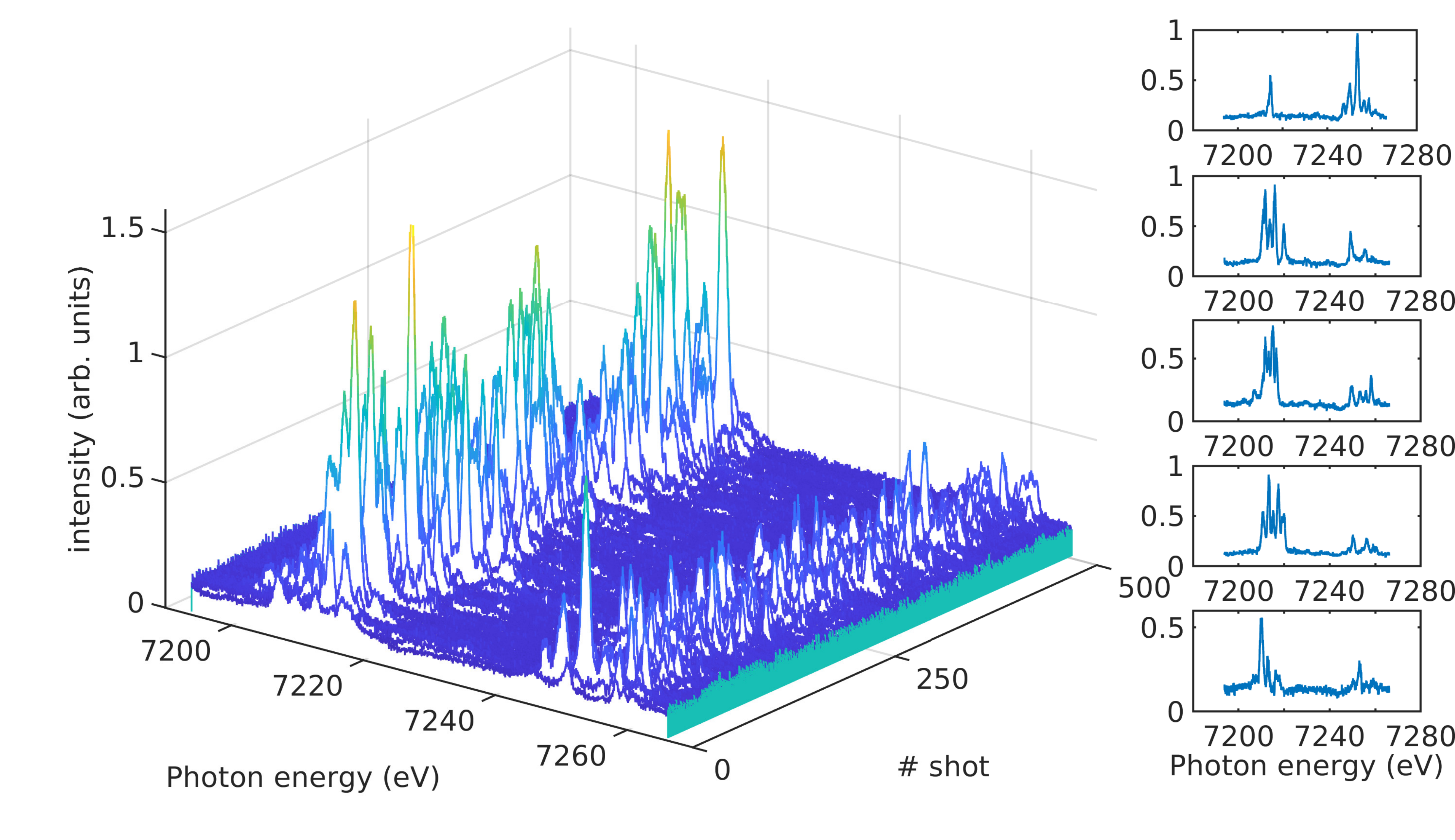}
	\caption{
		Single-shot spectra for 500 consecutive shots, showing the generation of two-color FEL pulses.
		A few individual spectra are displayed separately on the right.
	}
	\label{fig:psss}
\end{figure}

Figure~\ref{fig:psss} shows examples of single-shot two-color FEL spectra.
Here the beam tilt was generated with the second sextupole in BC1.
The bunch charge was 10~pC, fulfilling a requirement for unrelated beam studies that were performed priorly.
A moderate energy chirp was imposed onto the beam, so that the two pulses could be measured simultaneously in the PSSS.
In this particular case the intensity of the higher-energy pulse is generally lower compared to the other pulse.
The average photon pulse energy was on the order of 1~$\mu$J.
We estimate that the duration of each of the two pulses is at the fs level.
Beam-time constraints prevented us from realizing a more balanced spectrum and from obtaining higher pulse energies.
The integrated intensities of the two peaks are uncorrelated.
Most importantly, radiation from the core is fully suppressed.

For the measurements we present in the following, we used the large-bandwidth mode~\cite{Saa2016,Prat2020} to demonstrate a large energy separation between the two colors.
We employed a fixed compression setup in which the time and wavelength separation of the two pulses are linearly coupled.

The tunability of the method could be enhanced by using the longitudinal dispersion ($R_{56}$) of the energy collimator, as discussed in Ref.~\cite{Saa2019}.
For a given energy chirp, sextupole strength, and $c$ parameter, the wavelength separation of the two pulses would not change but the time difference would vary as a function of the $R_{56}$ and the energy of the two lasing slices.
Based on simulation studies presented in Ref.~\cite{Saa2019}, we estimate that the time separation of the two colors could be tuned within a maximum range between $-60$ and $+80$~fs.
Detailed studies on this matter are outside the scope of the paper.

SwissFEL operated with a bunch charge of 200~pC, a central electron beam energy of about 5.7~GeV, and a central photon energy of about 8750~eV.
We obtained an average pulse energy of 280~$\mu$J with an untilted beam.
A final energy chirp of about 1.5\%, a bunch length of about 65~fs, and an FEL bandwidth of about 1.7\% (all FWHM) were measured.
From the FEL resonance condition (Eq.~\ref{eq:resonance}) we deduce that about 57\% of the bunch contributes to the FWHM bandwidth, and estimate that 100~eV of energy separation correspond to about 25~fs of time separation.
The mean photon energy spectra were obtained with monochromator scans, since the full bandwidth of the produced radiation exceeded one PSSS window.
The resolution of the monochromator is less than 2~eV.
The monochromator step width was set between 15 and 25~eV.
At each step we recorded the intensity of 100 shots.
To every scan we apply a double Gaussian fit to extract the key properties of the measured spectra, i.e., peak intensities and widths of the two pulses, and the photon energy separation between them.

\begin{figure}[h]
	\centering
	\includegraphics[width=0.49\textwidth]{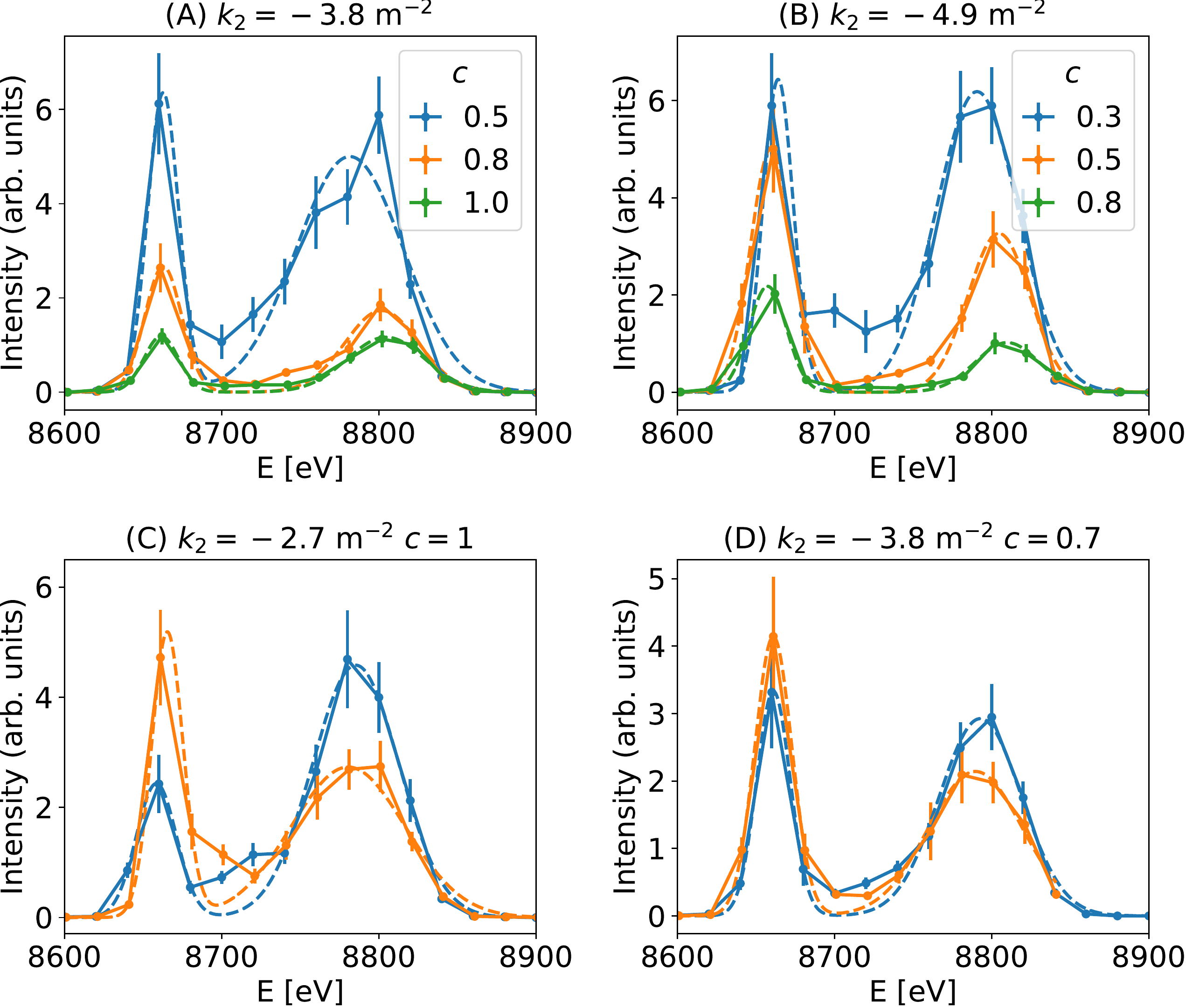}
	\caption{
		Monochromator scans showing the tunability of the method.
		A and B: different sextupole strengths ($k_2$) of the second sextupole in BC1 and orbit adjustments ($c$).
		C and D: two examples where the strength of a quadrupole in the EC is varied to balance the peak intensity of the two pulses (orange) with respect to the original cases (blue).
		}
	\label{fig:mono_scan}
\end{figure}

Figure~\ref{fig:mono_scan}~(A and B) shows monochromator scans of two-color FEL pulses.
Again the second sextupole in BC1 was used for beam tilt generation, with two different strengths.
In the plots we show the average spectral intensity, with error bars representing the estimated standard errors of the means.
A change in the orbit adjustment factor $c$ affects the FEL pulse as expected: when $c$ is increased, the radiation from the central slices is better suppressed, and the separation of the two lasing bunch slices increases.
The pulse intensity decreases substantially, which we attribute to the fact that the outer slices carry less beam current and the emittance is larger (as seen in Fig.~\ref{fig:m_scan}).
The widths of the lower-energy pulses are on the order of the photon energy jitter ($\approx$10~eV) and cannot be resolved with the multi-shot monochromator scan.
As expected, the widths of the higher-energy pulses are generally decreasing for higher $c$.
At larger sextupole strength a smaller $c$ parameter is sufficient to suppress radiation from the bunch core, again confirming our expectation.

In two-color operation modes, it is important to have control over the relative pulse energies of the two colors.
While the sextupole method in principle yields two symmetric pulses, we experienced that the balance of peak and integrated pulse intensities may be different.
We attribute the asymmetry to imperfections in the electron beam, such as an uncorrected linear beam tilt (as seen in Fig.~\ref{fig:m_scan}~A, although those measurements were done with a different machine setup).
Indeed, we found empirically that a superposition of a linear beam tilt, achieved by tweaking a quadrupole in the EC, can tune the balance between the two peaks.
Figure~\ref{fig:mono_scan}~(C and D) shows two examples of this simple adjustment with significant impact on the peak intensities of the two pulses.

\begin{figure}[h]
	\centering
	\includegraphics[width=0.48\textwidth]{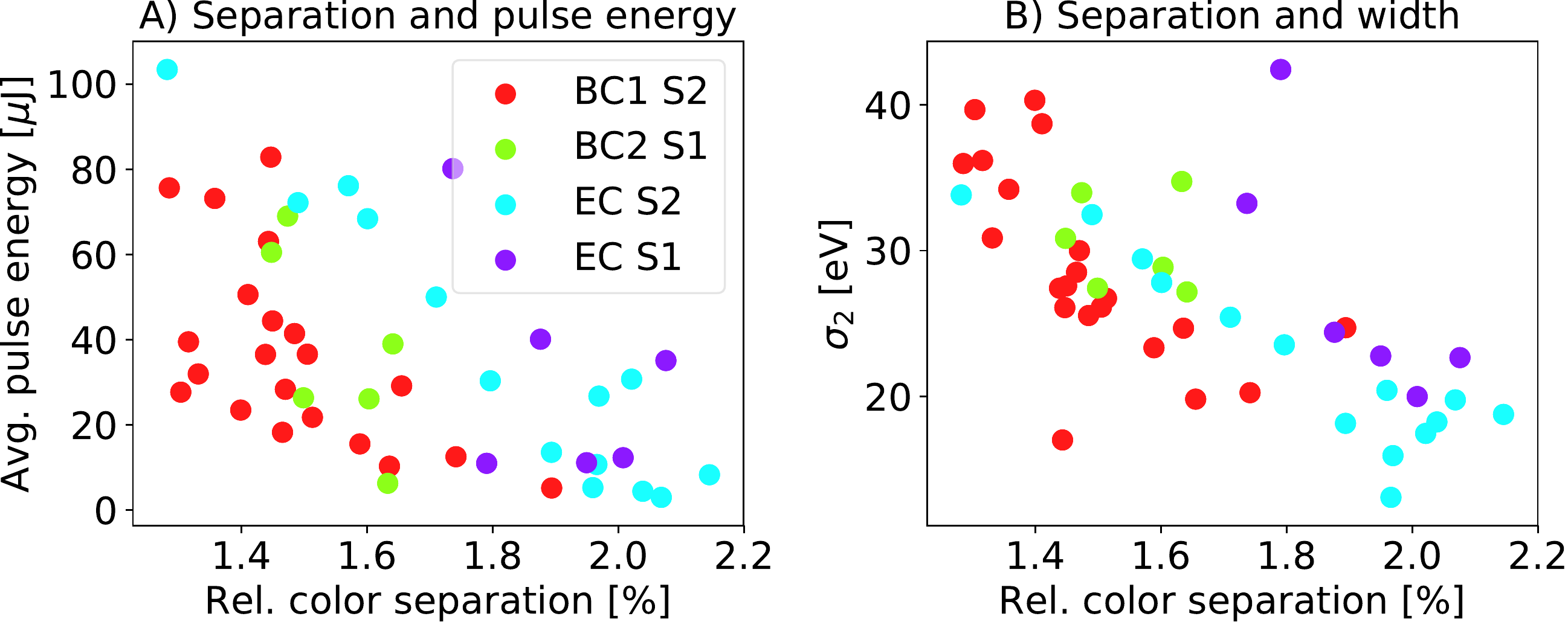}
	\caption{
		Properties of the two-color scheme measured with monochromator scans, while varying the strengths of four sextupoles and the orbit adjustment factor.
		A: color separation and average FEL pulse energy.
		B: color separation and width of the second pulse.
	}
	\label{fig:mono_scan3}
\end{figure}

An overview of pulse properties obtained in our experiment is provided in Fig.~\ref{fig:mono_scan3}.
Spectra were recorded while performing a 2D-scan of the sextupole strength and the orbit alignment factor $c$.
This was repeated for four different sextupoles.
We consider two-color pulses on the condition that radiation from the central part of the bunch be suppressed by at least a factor of three relative to the smaller of the two peaks.
Relative central photon energy separations between approximately 1.2\% and 2.2\% were observed,
corresponding to temporal separations between approximately 26 and 48~fs.
We assume the errors on the order of the monochromator step size of about 20~eV ($\approx$0.23\% relative separation).
The pulse energies reach values up to 100~$\mu$J, but are generally lower for the upper end of the photon energy separation.
In Fig.~\ref{fig:mono_scan3}~(B), the bandwidth of the second pulse is compared to the color separation.
We measured narrower pulses for larger spectral separations, as expected.

To conclude, we experimentally validated a new method of two-color FEL pulse generation.
Our approach requires only one sextupole magnet in a bunch compressor, is straightforward to set up and can work at any repetition rate (in contrast to other methods based on, e.g., an emittance spoiler~\cite{Saa2019}).
We are ready to apply the scheme to user experiments.

In principle, our method can be combined with two undulator sections tuned to different undulator strengths.
The beam alignment between the two sections can also be changed, so that fresh slices are used in the second undulator part.
As a result, three or even four photon pulses with different central photon energies are emitted from the same electron bunch.
We plan to investigate these options in future studies.

The raw data used for the Figs.~\ref{fig:m_scan}–\ref{fig:mono_scan3} is provided in Ref.~\cite{Dijkstal2020}

We acknowledge the technical groups involved in the operation of SwissFEL, in particular the laser, the diagnostics, the rf, and the operation teams.
We further thank Eugenio Ferrari and Simona Bettoni for participation in measurement shifts and valuable advice, Christopher Milne and Rolf Follath for their help in setting up the monochromator, and Thomas Schietinger for editing the language of the manuscript.
This work was supported by the SNF grant 200021\_175498.

\bibliography{./Phd_thesis.bib}
\end{document}